\documentclass[intlimits,twoside,a4paper]{article}

\usepackage{amsmath,amssymb}
\usepackage{graphicx}
\usepackage[T2A]{fontenc}
\usepackage[cp1251]{inputenc}

\usepackage{textalpha}
\usepackage[eqsecnum]{cmpj2}

 \def\coppa{{\fontencoding{LGR}\fontfamily{cmr}\selectfont\textqoppa}}
 \def\qq{\textrm{\coppa}}
 \def\sqq{\textrm{\coppa}}

\issue{2017}{20}{1}{13601}
\doinumber{10.5488/CMP.20.13601}

\title[Revisiting (logarithmic) scaling relations]%
{Revisiting (logarithmic) scaling relations using renormalization
   group}
\author{J.J. Ruiz-Lorenzo\refaddr{label1,label2,label3}}
\addresses{
  \addr{label1} Departamento de F\'{\i}sica, Universidad de Extremadura, 06071
  Badajoz, Spain  \addr{label2} Instituto de Computaci\'on Cient\'{\i}fica
  Avanzada (ICCAEx), Universidad de Extremadura, 06071 Badajoz, Spain
  \addr{label3} Instituto de Biocomputaci\'on y F\'{\i}sica de los Sistemas
  Complejos (BIFI), Zaragoza, Spain}

\authorcopyright{J.J. Ruiz-Lorenzo, 2017}
\date{Received January 11, 2017, in final form January 23, 2017}

\begin{document}

\maketitle

\begin{abstract}

We explicitly compute the critical exponents associated with logarithmic
corrections (the so-called hatted exponents) starting from the renormalization
group equations and the mean field behavior for a wide class of models at the
upper critical behavior (for short and long range $\phi^n$-theories) and below
it. This allows us to check the scaling relations among these critical
exponents obtained by analysing the complex singularities (Lee-Yang and Fisher
zeroes) of these models. Moreover, we have obtained an explicit method to
compute the $\hat{\qq}$ exponent [defined by $\xi\sim L (\log L)^{\hat{\qq}}$]
and, finally, we have found a new derivation of the scaling law associated
with it.

\pacs 64.60-j,05.50+q,05.70.Jk,75.10.Hk
\end{abstract}

\section{Introduction}

One of the main achievements of Wilson's~\cite{KGW} renormalization group (RG)
was the definition of universality class by means of a finite number of
critical exponents. These critical exponents determine the divergences of some
observables at the critical point \cite{Parisi,Amit,Cardy,Itzykson,LeBellac}.

In particular circumstances, logarithmic corrections arise multiplicatively
in these critical laws. These logarithms are of  a paramount importance in some
materials (for example, dipolar magnets in three dimensions, which is the
upper critical dimension of the system \cite{AAA}) and can be accessed
experimentally~\cite{Exp}. Moreover, their effects are very important in the non-perturbative
definition of quantum field theories in four dimensions (the so-called
triviality problem) \cite{Sokal}.

In \cite{RalphA,RalphB,RalphD}, the scaling relations of the
exponents which characterize the logarithmic corrections were derived using
the Lee-Yang~\cite{LY} and Fisher zeroes~\cite{FZ} techniques in a
model-independent manner. In this paper, we will explicitly compute, using RG and field
theory, the value of these exponents and then check the (scaling)
relations among them. We have done this for a wide class of models [$\phi^n$
models at their upper critical dimensions with short (SR) and long range (LR)
interactions] and can also be applied to the models in low dimensions (as the
four-state Potts model in two dimensions).

In the presence of logarithmic corrections, the scaling laws for the observables
near the critical point must be modified
as~\cite{Parisi,Amit,Cardy,Itzykson,LeBellac}\footnote{We use in the definition of the critical
  exponents the standard notation, see, for example, \cite{Parisi,Amit,Cardy,Itzykson,LeBellac,RalphA,RalphB}.}
\begin{eqnarray}
  \label{eq:xi}
\xi&\sim & |t|^{-\nu} \big|\log |t|\big|^{\hat \nu}\,, \\
\label{eq:C}
  C&\sim & |t|^{-\alpha} \big|\log |t|\big|^{\hat \alpha}\,, \\
  \label{eq:m}
  m&\sim & |t|^\beta \big|\log |t|\big|^{\hat \beta}\,\,\,\text{for} \,\,\,t<0\,,\\
  \label{eq:chi}
\chi&\sim & |t|^{-\gamma} \big|\log |t|\big|^{\hat \gamma}\,,
\end{eqnarray}
\vspace{-6mm}
\begin{eqnarray}
    \label{eq:mh}
  m&\sim & h^{1/\delta} |\log h|^{\hat \delta}\,\,\,\text{for} \,\,\,t=0\,,\\
  \label{eq:LY}
  r_\text{LY}&\sim & |t|^\Delta \big|\log |t|\big| ^{\hat \Delta}\,,\\
\label{eq:gr}
G(r)&\sim & \frac{(\log r) ^{\hat \eta}}{r^{d-2 +\eta}}  \,\,\,\text{for} \,\,\,t=0\,,
\end{eqnarray}
which define the so-called hatted exponents ($d$ being the dimension).  The standard critical exponents
(e.g., $\alpha$, $\beta$, $\gamma$, etc.) satisfy the classic scaling laws (see,
for example, \cite{Parisi,Amit,Cardy,Itzykson,LeBellac}). In addition, in \cite{RalphA,RalphB,RalphD}, it was shown that the hatted exponents satisfy the following
scaling relations:\footnote{Recall $\delta=(d+2-\eta)/(d-2+\eta)$.}
\begin{eqnarray}
        \label{eq:bgd}
        \hat{\Delta}&=& \hat{\beta}-\hat{\gamma}\,, \\
        \label{eq:bdg}
        \hat{\beta}(\delta-1)&=&\delta \hat{\delta}-\hat{\gamma}\,,\\
        \label{eq:eta}
        \hat{\eta}&=&\hat{\gamma}-{\hat \nu} (2-\eta)\,.
\end{eqnarray}
Finally, see \cite{RalphA,RalphB,RalphD}, at the infinite volume critical point, the
correlation length of the system defined on a finite box of size $L$ behaves
as\footnote{In this paper we avoid the mean field region by working at and
  below the upper critical region.}
\begin{equation}
\label{eq:q}
\xi \sim L (\log L)^{\hat{\qq}}
\end{equation}
and the associated related scaling relation is
\begin{equation}
  \label{eq:qscaling}
{\hat \alpha}= d {\hat{\qq}} -d {\hat \nu}\,.
\end{equation}
When $\alpha=0$ and the impact angle of the Fisher zeros satisfies $\phi\neq
\pi/4$, the previous relation should be  modified as~\cite{RalphB}
\begin{equation}
  \label{eq:qscalingmod}
{\hat \alpha}= 1+ d {\hat{\qq}} -d {\hat \nu}\,.
\end{equation}
Finally, an additional scaling relation can be written~\cite{RalphD}
\begin{equation}
  \label{eq:abg}
 2 {\hat \beta}-{\hat \gamma}= d {\hat{\qq}} -d {\hat \nu}\,. \\
\end{equation}

In this paper we will mainly analyze generic $\phi^n$ theories, with
Hamiltonian (for simplicity we write the scalar version for the short range
model):\footnote{Using power counting, we can compute when the coupling $g_n$
  is marginal, obtaining the so-called upper critical dimension, that for
  short range models is
$$d_u=\frac{2 n}{n-2}$$
and for long range models (with propagator $1/q^\sigma$)
$$d_u=\frac{n \sigma}{n-2}\,.$$
For $\sigma=2$, we recover the short range result.
}
\begin{equation}
{\cal H}=\int \rd^d x \left[\frac{1}{2} (\partial_\mu \phi)^2 + \frac{1}{2} r_0
\phi^2 + \frac{1}{n} g_n \phi^n \right]\,.
\end{equation}
\section{Some mean field results}
We will use RG to analyze the critical behavior of the models, and after a
finite number of RG step we will finish in the parameter region in which we
can apply mean field results. In this section we will briefly review the basic
facts of the scaling in this mean field region \cite{Cardy,Itzykson}.

We start with the free energy per spin for a $\phi^n$-theory:
\begin{equation}
f(m)=\frac{r_0}{2} m^2+ \frac{g_n}{n} m^n\,.
\label{eq:mf}
\end{equation}
Minimizing $f(m)$, for $r_0<0$, we obtain magnetization as:
\begin{equation}
m=\left(\frac{|r_0|}{g_n}\right)^{1/(n-2)}\sim \frac{1}{g_n^{p_m}}\,,
\label{eq:M}
\end{equation}
where $p_m=1/(n-2)$ and
\begin{equation}
f_\text{min}\propto \frac{r_0^{n/(n-2)}}{g_n^{2/(n-2)}}\,.
\end{equation}
The susceptibility is
\begin{equation}
\chi \propto |r_0| \,,
\label{eq:Chi}
\end{equation}
and the specific heat
\begin{equation}
C \propto \frac{r_0^{2/(n-2)}}{g_n^{2/(n-2)}} \sim \frac{1}{g_n^{p_c}}\,,
\label{eq:Cmf}
\end{equation}
where $p_c=2/(n-2)$.
Finally, we can add a magnetic field [which induces a term $- h m$ in
equation  (\ref{eq:mf})] and compute the minimum of the free energy just at the
critical point, $r_0=0$
(which is relevant in the computation of the critical isotherm)
\begin{equation}
f_\text{min}(r_0=0, h) \propto  \frac{h^{n/(n-1)}}{g_n^{1/(n-1)}}\,,
\label{eq:D}
\end{equation}
and the magnetization at criticality is
\begin{equation}
m(r_0=0, h) \propto  \left(\frac{h}{g_n}\right)^{p_h}\,,
\label{eq:Dh}
\end{equation}
where $p_h=1/(n-1)$.\footnote{The introduction  of $p_m$, $p_h$ and $p_c$ will
  be useful at the upper critical dimension to collect the extra logs yielded
  by the $g$ renormalizing to zero in a logarithmic way. Below the upper critical dimension, $g_n$ is not a
  dangerous irrelevant variable: in
  this situation, we will use $p_m=p_c=p_h=0$, i.e., there will be no extra logs
  from the $g_n(b)$ in the mean field region.}
Hence, since $n>2$, $g_n$ is an irrelevant  dangerous variable for
magnetization, critical isotherm  and
 specific heat, yet, $\chi$ is free of this problem.

\section{Revisiting logarithmic corrections}
\label{sec:tres}

The starting point is the behavior of  the singular part of the free energy
density (that we denote simply as $f$ and denoting $g_n$ by $g$) under a RG
transformation
\begin{equation}
f(t_0, h_0, g_0)=\frac{1}{b^d}f[t(b), g(b), h(b)] \,,
\label{eq:f}
\end{equation}
where $b$ is the RG scaling factor and  $t(b)$, $h(b)$ and $g(b)$ (the running
couplings) denote the evolution of different
couplings under a RG transformation, which are
obtained solving the following differential equations (we write them for the
LR model)
\begin{eqnarray}
\frac{\rd t }{\rd \log b}&=& t[\sigma+\overline{\gamma}(g)]\,,\\
\frac{\rd \log h }{\rd \log b}&=&\frac{d}{2}+1 -\frac{\gamma}{2}\,,\\
\frac{\rd g }{\rd \log b}&=& \beta_\text{W}(g)\,,
\end{eqnarray}
which define the functions $\beta_\text{W}, \gamma$ and
$\overline{\gamma}$.\footnote{We can compute the thermal and magnetic critical
 exponents by means of $\eta=\gamma(g^*)$ and $1/\nu=\sigma+\overline{\gamma}(g^*)$,
where $g^*$ satisfies $\beta_\text{w}(g^*)=0$ \cite{Amit,JJA}.}
For further use we define two functions $F(b)$ and $\zeta(b)$ and we assume the
following asymptotic behavior [$g_0\equiv g(1)$]
\begin{eqnarray}
F(g)&\equiv&\exp\left[\int^{g(b)}_{g_0} \rd g
  \frac{\overline{\gamma}(g)}{\beta_\text{W}(g)}\right]\sim b^a (\log b)^p\,,\\
\zeta(g)&\equiv& \exp\left[-\frac{1}{2}\int^{g(b)}_{g_0} \rd g
  \frac{\gamma(g)}{\beta_\text{W}(g)}\right]\sim b^c (\log b)^x \,.
\end{eqnarray}
The solutions are (we also write the asymptotic behavior as $b\to \infty$) as follows:
\begin{eqnarray}
\label{eq:t}
t(b)&=&t_0 b^\sigma \exp\left[\int^{g(b)}_{g_0} \rd g
  \frac{\overline{\gamma}(g)}{\beta_\text{W}(g)}\right]\sim t_0 b^{\sigma+a} (\log b)^{p}\,,\\
\label{eq:h}
h(b)&=&h_0 b^{\frac{d}{2}+1} \exp\left[-\frac{1}{2}\int^{g(b)}_{g_0} \rd g
  \frac{\gamma(g)}{\beta_\text{W}(g)}\right]\sim
h_0 b^{\frac{d}{2}+1+c} (\log b)^x\,,\\
\log b&=& \exp\left[\int^{g(b)}_{g_0} \rd g
  \frac{1}{\beta_\text{W}(g)}\right] \,.
\label{eq:g}
\end{eqnarray}
In the asymptotic regime (and for the models under consideration in this paper
where $\beta_\text{W} \propto g^s$),
the last equation can be written as
\begin{equation}
g(b) \sim (\log b)^{-r} \,,
\label{eq:gren}
\end{equation}
and this defines the $r$ exponent ($1/r=s-1$).  In particular, the useful relation
$t(b^*)=1$ can be written as
\begin{equation}
\label{33}
 b^*\sim t_0^{-1/(\sigma+a)} (\log t_0)^{-p/(\sigma+a)}\,.
 \end{equation} 
Therefore, we can identify $\nu=1/(\sigma+a)$ and $\hat{\nu}=-p/(\sigma+a)=-p
\nu$. From the form of $h(b)$, one can obtain $c=-\eta/2$. By computing
suitable derivatives of the free energy per spin [see equation~(\ref{eq:f})]
and using the renormalized couplings given by equations~(\ref{eq:t})--(\ref{eq:g}), and in the case of the upper critical
dimension using the expression of the intensive free energy in the mean field
regime [equations (\ref{eq:M}), (\ref{eq:Chi})--(\ref{eq:D})], we can
obtain the following relations for the exponents which control the logarithmic
corrections (see the appendix for more details)
\begin{equation}
\hat{\alpha}=-d \hat{\nu} + r p_c \,,
\label{eq:alpha}
\end{equation}

\begin{equation}
\hat{\gamma}=\hat{\nu} (2-\eta)+ 2 x \,,
\label{eq:gammaC}
\end{equation}

\begin{equation}
\hat{\beta}=-\hat{\nu} \left(\frac{d}{2}-1 +\frac{\eta}{2}\right)+x+r p_m \,,
\label{eq:betaC}
\end{equation}

\begin{equation}
\hat{\delta}=\frac{2 x d}{d+2-\eta} + r p_h \,,
\label{eq:deltaC}
\end{equation}

\begin{equation}
\hat{\Delta}= -\hat{\nu} \left(\frac{d}{2}+1 -\frac{\eta}{2}\right)-x + r p_m \,,
\label{eq:DeltaC}
\end{equation}

\begin{equation}
\hat{\eta}= 2 x \,.
\label{eq:etaC}
\end{equation}

These equation must be read at the upper critical dimension with $\eta=0$ (SR)
or $\eta=2-\sigma$ (LR) and $d=d_u$, otherwise, below $d_u$, all the $p$'s
from the mean field are zero ($p_m=p_c=p_h=0$).  With these explicit expressions
for the hatted exponents, it is easy to re-derive the scaling relations given
by equations~(\ref{eq:bgd})--(\ref{eq:eta}), (\ref{eq:abg}).

In models with  $\alpha=0$ and impact angle of the Fisher zeroes $\phi\neq
\pi/4$, a circumstance equivalent to $A_{-}/A_{+} = 1$ (being $A_{\pm}$ the
critical amplitudes of the specific heat)~\cite{SalasSokalB},
the scaling of the free energy is modified as\footnote{As described in
  \cite{SalasSokalB} the appearance of this
  extra log term in the free energy can be explained either as a resonance between
  the thermal and the identity operators or as an interplay between
  the singular and regular parts of the free energy.}
\begin{equation}
f(t_0, h_0, g_0)=\frac{1}{b^d}f\big(t(b), g(b), h(b)\big) +
\frac{1}{b^d} (\log b) f_l\big(t(b), g(b), h(b)\big)\,,
\label{eq:fmd}
\end{equation}
where the functions $f$ and $f_l$ satisfy additional constraints to generate
the right logarithmic corrections (for more details see \cite{SalasSokalB} and references therein). This decomposition of the free
energy can be also understood in terms of a Lee-Yang and Fisher zeros analysis,
see \cite{RalphB,RalphD}.  For instance, in the two-dimensional
pure Ising model, only the ``energy''-sector develops logarithmic corrections,
and these corrections (for the free energy, energy and specific heat) are
provided by the term proportional to $f_l$. However, the scaling of the
``magnetic''-sector is given by the standard term, proportional to $f$. In the
two dimensional diluted Ising model, the magnetic sector also shows logarithmic
corrections, provided by the (standard) term proportional to $f$, whereas the
corrections for the energy-sector are given by the term proportional to
$f_l$. Hence, only the relation of $\hat \alpha$ (which is computed with the
$f_l$-term) should be modified
\begin{equation}
\hat{\alpha}=-d \hat{\nu} + r p_c +1\,.
\label{eq:alpham}
\end{equation}

We have checked that these equations provide correct hatted exponents in
$O(N)$-$\phi^4$ models\footnote{Where $N$ is the number of components of the
  field.}  in the short range and long range interactions, tensor (short
range) $\phi^3$ (which includes percolation, $m$-compo\-nent spin glasses and
Lee-Yang singularities, and can also be related with lattice animals), all of
them at their upper critical dimension and in the four-state Potts models,
pure Ising model and diluted Ising model in two
dimensions \cite{JJA,LR,RalphC,RalphD,SalasSokal,Shchur}.  The logarithmic scaling
relations for all these models were thoroughly checked in
\cite{RalphD}.\footnote{In \cite{RalphD} other exponents
  were defined (e.g., ${\hat \epsilon}$, ${\hat \nu_c}$ and ${\hat
    \alpha_c}$). It is straightforward to compute them using the theoretical
  framework of this paper.}

Finally, using this theoretical framework we have been able to compute $\hat
\Delta$ for the four-state two-dimensional Potts model, $\hat \Delta$, $\hat
\beta$, $\hat \eta$ and $\hat \delta$ for SR tensor $\phi^3$-theories and
$\hat \Delta$ for the LR $O(N)$ $\phi^4$-theories. Finally, $\hat{\qq}$ has
been computed for the LR $O(N)$ $\phi^4$-theories. The numerical values for
all these exponents were derived in references \cite{RalphA,RalphB,RalphD}
using the logarithmic scaling relations
(\ref{eq:bgd})--(\ref{eq:eta}), (\ref{eq:qscaling}), (\ref{eq:abg}). See
\cite{RalphD} for the values of these hatted exponents.

\section{A re-derivation of ${\hat \alpha}= d {\hat{\sqq}} -d {\hat \nu}$}
\label{sec:q}

We start with the dependence of a singular part of the intensive free energy on $L$
\begin{equation}
\label{42}
f_\text{sing}\propto L^{-d} \,.
\end{equation}
This is the key point of the derivation. Below the upper critical dimension,
one has $L\sim \xi$ and one can write $f_\text{sing}\propto \xi^{-d}$, but due
to the logarithmic corrections which appear at the upper critical dimension
this is no longer true.

We can also write the singular part of the free
energy, using the scaling of the specific heat [see equation  (\ref{eq:C})], as
\begin{equation}
\label{43}
f_\text{sing}\propto L^{-d}\propto t^{2-\alpha} (\log t)^{\hat \alpha} \,.
\end{equation}
Using equations (\ref{eq:xi}) and (\ref{eq:q}) one can write
\begin{equation}
L^{-d} \sim \xi^{-d} (\log \xi)^{d {\hat{\qq}}}\sim t^{\nu d} (\log t)^{d {\hat{\qq}} -d {\hat \nu}} \sim t^{2-\alpha} (\log t) ^{\hat \alpha} \,.
\end{equation}
Identifying the exponents of $\log t$ of the last two expressions we obtain the
scaling relation given by equation  ({\ref{eq:qscaling}).

When $\alpha=0$ and $\phi\neq \pi/4$ \cite{RalphB}, the free
energy scales as $f\propto L^{-d} \log L$ [see equation (\ref{eq:fmd}) and the
discussion of section \ref{sec:tres}]. This extra-log, using the previous
arguments, provides the following scaling law:
\begin{equation}
{\hat \alpha}= 1 + d ({\hat{\qq}}- {\hat  \nu})\,,
\end{equation}
obtaining equation ({\ref{eq:qscalingmod}).

\section{Computation of the $\hat{\sqq}$-exponent}

We will compute the exponent $\hat{\qq}$ for a generic
 $\phi^n$ theory at its upper
critical dimension for both short and long range models. The starting
point is the expression of $\chi$ in terms of the free energy\footnote{Since, in this section, we work with the
 susceptibility, we take into account only the term
  proportional to $f$ in equation~(\ref{eq:fmd}) independently of the value of
  $\alpha$ and $\phi$. See discussion of section \ref{sec:tres}.}
\begin{equation}
\chi\sim b^2 \zeta^2 \left.\frac{\partial^2 f\big(t(b),g(b),h\big)}{\partial h^2}\right\arrowvert_{h=0}\,.
\end{equation}
This can be written as [using $t(b^*)=1$ and $b^* \sim \xi$]
\begin{equation}
\chi \sim  \zeta(\xi)^2 \xi^2 \propto \xi^{2+2 c} (\log \xi) ^{2 x}\,.
\label{eq:xi_xeta_xi}
\end{equation}

In a $\phi^n$ theory we can rescale the field via $\phi^\prime =g^{1/n} \phi$~\cite{Luitjen},
and the free energy per spin verifies
\begin{equation}
 f(t_0, g_0 ,h_0)=L^{-d}
 G\left(\frac{t(L)}{g^{2/n}}\,,\frac{h(L)}{g^{1/n}}\right) \,.
 \label{eqFL}
 \end{equation} 
Differentiating twice equation  (\ref{eqFL}) with respect to the magnetic field ($h_0$), we obtain
\begin{equation}
 \chi\propto L^{-d} \left[\frac{\partial h(L)}{\partial h_0}  \right]^2
 \left.\frac{\partial^2}{\partial h(L)^2}
 G\left(\frac{t(L)}{g(L)^{2/n}}\,,
 \frac{h(L)}{g(L)^{1/n}}\right)\right\arrowvert_{h_0=t_0=0}\sim L^2
 \zeta(L)^2 \frac{1}{g(L)^{2/n}} \,.
 \end{equation} 
Comparing with equation  (\ref{eq:xi_xeta_xi}), we finally obtain
\begin{equation}
\xi \sim \frac{L}{g(L)^{2/[n(2+2 c)]}} \,,
\end{equation}
and assuming the asymptotic behavior of $g(L)$ given in equation  ({\ref{eq:gren}) we  finally get
\begin{equation}
\label{eq:qexplicit}
{\hat{\qq}}=\frac{2r}{n(2+2c)}=\frac{2r}{n(2-\eta)}\,.
\end{equation}
For the short range $\phi^4$ theory ($\sigma=2$, $n=4$, $\eta=-2 c=0$ and $r=1$) we obtain
$\hat{\qq}=1/4$. For the short range $\phi^3$ theory ($\sigma=2$, $n=3$, $\eta=-2 c=0$ and
$r=1/2$) we get ${\hat{\qq}}=1/6$. In addition, for the long range $\phi^4$
model [$n=4$, $\eta=-2 c=(2-\sigma)$ and $r=1$], ${\hat{\qq}}=1/(2 \sigma)$.

Another way to obtain $\hat{\qq}$ is to use the scaling relation provided by
equation (\ref{eq:qscaling}) and equation (\ref{eq:alpha})
\begin{equation}
\label{eq:hatq}
\hat{\qq}=\frac{\hat{\alpha}}{d}+\hat{\nu}=\frac{r p_c}{d}
\end{equation}
or for $\alpha=0$
and $\phi\neq \pi/4$, equations~(\ref{eq:qscalingmod}), (\ref{eq:alpham})
\begin{equation}
\label{eq:hatqm}
\hat{\qq}=\frac{\hat{\alpha}}{d}+\hat{\nu}-\frac{1}{d}=\frac{r p_c}{d}\,,
\end{equation}
obtaining the same final result irrespectively of the value of $\alpha$ and
the impact angle $\phi$.

So, $\hat{\qq}=0$ below $d_u$ since $p_c=0$ therein; at the
upper critical dimension (SR models) $d=d_u=2 n/(n-2)$, then $\hat{\qq}=r/n$
as computed before. For LR models, $d_u= n \sigma/(n-2)$ and then we recover
the result given by equation (\ref{eq:qexplicit}).

In \cite{JJA,JJB}, the $\hat{\qq}$ exponent was computed using a
misidentification of the correlation length for a lattice of size $L=1$ [see
equations (\ref{42}), (\ref{43}) and (\ref{33}), (\ref{eq:alpha}) of \cite{JJB} and \cite{JJA},
respectively], providing, however, with the correct value of ${\hat \qq}$ in
general $\phi^4$ theories (and, in particular, for the four dimensional
diluted model, see reference~\cite{JJB}, where the right value of ${\hat
  \qq}=1/8$ was obtained~\cite{JJB}) but not in $\phi^3$ ones~\cite{JJA}. In
this section we have developed a new general method which avoids the previous
misidentification of $\xi$.  In particular, we have obtained the correct value of
${\hat \qq}=1/6$ for the general class of $\phi^3$ theories, see above.

In \cite{RKBB} it was conjectured that there is a relationship between
$\hat{\qq}$ and $1/d_u$ which is frequently an equality but not always so.
Indeed, It was already known~\cite{AGG} that $\hat{\qq}=1/8$ for the four
dimensional Ising model which is described by a $\phi^4$ theory which has
$d_u=4$. In this paper we have provided the general relation between
$\hat{\qq}$ and $d_u$.

To finish this section, we present two examples in which $\hat{\qq} \neq
1/d_u$ to understand the reasons behind the modification of this behavior. The
first one is based on the study of $\phi^{2 k}$-theories with $k>2$ and the second one is
the two parameter $\phi^4$-theory which describes the four dimensional diluted
Ising model.

\subsection{$\phi^{2 k}$-theories with $k>2$ and short range interactions}

The upper critical dimension for these models is $d_u=2 k/(k-1)$. One can
compute the RG equations at $d_u$ obtaining~\cite{Itzykson}
\begin{equation}
 \frac{\rd g_{2 k}}{\rd \log b}\propto g_{2 k}^2
\end{equation}
and so $g_{2 k}\propto 1/\log L$ ($r=1$), that  using equation  (\ref{eq:qexplicit}) provides
${\hat{\qq}}=1/(2 k)$ which is different to
${\hat{\qq}}=(k-1)/(2 k)$ (only works for $k=2$).\footnote{In addition,
  working at $d_u$ for short range models, $\eta=0$ and so $c=0$.}

\subsection{Diluted Ising model}
One can obtain an effective field theoretical version of the diluted Ising model
by using the replica trick, with effective Hamiltonian given by~\cite{JJB,AAB}
\begin{equation}
\label{action}
{\cal H}_{\mathrm{eff}}[\phi_i]=\!\int\! \rd^d x \left[ \frac{1}{2} \sum_{i=1}^n
\left(\partial_\mu \phi_i\right)^2 +
\frac{r}{2} \sum_{i=1}^n \phi_i^2 +\frac{u}{4!} \left(\sum_{i=1}^n
 \phi_i^2\right)^2 + \frac{v}{4!} \sum_{i=1}^n  \phi_i^4 \right]\!,\!
\end{equation}
where $v$ is related with the original Ising coupling and $u$ is a function of
the disorder strength. In the replica trick it is mandatory to take the limit
of the number of replicas, $n$, to zero ($i=1,\dots,n$). The RG equations are, in $d=4$ and $n=0$,
\begin{eqnarray}
\frac{\rd r}{\rd \log b} &= & 2 r+4  (2u+3 v) (1-r) \ ,\\
\label{rg_eq}
\frac{\rd v}{\rd \log b} &= &- 12  v (4 u + 3v) \ , \\
\frac{\rd u}{\rd \log b} &=& - 8  u (4 u + 3 v) \ .
\end{eqnarray}
In the standard
$\phi^4$ theory one gets $\beta \propto g^2$. Hence,
$g\propto 1/\log L$ and ${\hat{\qq}}=1/4$. However,  the RG flow of the diluted model
 asymptotically finishes on the line $4 u + 3 v =O(u^2)$, so we
need to include the next (cubic) terms in the perturbative expression and
the RG $\beta$-functions are no longer quadratic in the couplings. Finally, one finds
that $u(b)^2 \sim v(b)^2 \sim 1/\log b$: hence, ${\hat{\qq}}=1/8$ as derived
in \cite{JJB,AGG}.

\section{Conclusions}

By explicitly computing the hatted critical exponents for a wide family of
models we have been able to check the scaling relations among them using the
RG framework and the behavior in the mean field regime.  Some of these hatted
exponents (for some of the models) have been previously derived by using the
logarithm scaling relations.

In addition, we have generalized a conjecture regarding a relationship between
${\hat\qq}$ and $d_u$ and derived~it.

Finally, we have found a new method to derive the scaling relation
associated with $\hat{\qq}$ and we have briefly discussed the logarithmic
corrections to the free energy when the Fisher zeros have an impact angle
other than $\pi/4$ and $\alpha=0$.

\section*{Acknowledgements}

I dedicate this paper to Y. Holovatch to celebrate his 60th birthday.

I acknowledge interesting discussions with R. Kenna, B. Berche and M. Dudka.
This work was partially supported by Ministerio de Econom\'{\i}a y
Competitividad (Spain) through Grants No.\ FIS2013-42840-P and
FIS2016-76359-P (partially funded by FEDER) and  by Junta de Extremadura (Spain) through Grant No.\ GRU10158
(partially funded by FEDER).

\appendix
\section{Appendix}

In this appendix we give additional details of the computation of the
hatted exponents, see section \ref{sec:tres}.

By differentiating once the free energy (\ref{eq:f}) with respect
to the magnetic field, then renormalizing to $t(b^*)=1$, and finally evaluating
the magnetization using the mean field behavior (\ref{eq:M}), we obtain
\begin{equation}
m \propto (b^*)^{-d} (b^*)^{d/2+1}  \exp\left[-\frac{1}{2}\int_{g_0}^{g(b^*)}
\rd g ~\frac{ \gamma(g)}{\beta_{\text W}(g)} \right] \frac{1}{g(b^*)^{p_m}}\sim
(b^*)^{-d/2+1+c} (\log b^*)^{x+p_m r} \, .
\label{eq:mA}
\end{equation}
The susceptibility is obtained by differentiating twice the free energy with
respect to the magnetic field [notice that there is no dependence on $g$ in
the mean field region  (\ref{eq:Chi})]:
\begin{equation}
\chi\propto (b^*)^{-d} (b^*)^{d+2} \exp\left[-\int_{g_0}^{g(b^*)}
\rd g ~\frac{ \gamma(g)}{\beta_{\text W}(g)} \right] \sim
(b^*)^{2+2 c} (\log b^*)^{2 x}\, .
\label{eq:chiA}
\end{equation}
To obtain the specific heat, we differentiate twice the free energy  with respect to the
temperature, renormalize to $t(b^*)=1$,  and evaluate the specific heat using the
mean field behavior (\ref{eq:C}), obtaining
\begin{equation}
C \propto (b^*)^{-d} (b^*)^{2 \sigma}  \exp\left[2 \int_{g_0}^{g(b^*)}
 \rd g~\frac{ {\overline{\gamma}(g)}}{\beta_{\text W}(g)} \right]
\frac{1}{g(b^*)^{p_c}} \sim
(b^*)^{-d+ 2 \sigma+ 2 a} (\log b^*)^{2 p+r p_c}
\, .
\label{eq:CA}
\end{equation}
The correlation length is obtained from $t(b^*) =1$
\begin{equation}
\xi \propto b^* \sim t_0^{-1/(\sigma+a)} (\log t_0)^{-p/(\sigma+a)} \, .
\label{eq:xiA}
\end{equation}
By putting the previous relation between $b^*$ and $t_0$ in equations
(\ref{eq:mA})--(\ref{eq:CA}) and matching the l.h.s. logarithms
[given by equations (\ref{eq:C})--(\ref{eq:chi})] with
the r.h.s. ones [given by equations (\ref{eq:mA})--(\ref{eq:CA})] we obtain equations~(\ref{eq:alpha})--(\ref{eq:betaC}).

To compute the Lee-Yang edge, the starting point is the renormalized potential~\cite{Itzykson}
\begin{equation}
V\big(t(b), g(b), h(b)\big)=\frac{t(b)}{2} m^2+
  \frac{g(b)}{n} m^n -h(b) m\,.
\end{equation}
From the constraints $\partial V/\partial m=0$ and $\partial^2
V/\partial m^2=0$ and working in the broken phase with $t(b^*) =-1$, it is possible to show that
\begin{equation}
h(b^*) \sim m(b^*) \sim 1/g(b^*)^{p_m} \,,
\end{equation}
that can be written as
\begin{equation}
h(b^*)=h_0 (b^*)^{\frac{d}{2}+1} \exp\left[-\frac{1}{2}\int^{g(b^*)}_{g_0} \rd g
  \frac{\gamma(g)}{\beta_\text{W}(g)}\right]\sim
h_0 (b^*)^{\frac{d}{2}+1+c} (\log b^*)^x \sim (\log b^*)^{r p_m} \,,
\end{equation}
which allows us to compute $h_0$ as a function of $b^*$, and knowing
$b^*(t_0)$ (\ref{eq:xiA}), we can easily obtain  $h_0(t_0)$. The comparison of  the logarithm of
$h_0(t_0)$ with that of equation (\ref{eq:LY}) provides us with relation (\ref{eq:DeltaC}).

Relation (\ref{eq:etaC}) can be obtained taking the Fourier transform of
equation (\ref{eq:gr}) at $b\sim 1/q$ ($q$ being the momentum), and comparing
this with the renormalized propagator in momentum space (see \cite{LeBellac}).

Finally, for the critical isotherm, we start with the free energy computed at the
critical point \linebreak $f(0,h_0, g_0)$, differentiate once with respect to $h_0$ to
compute the critical magnetization,
then renormalize to  $h(b^*)=1$ and use the  mean field behavior of the
critical magnetization (\ref{eq:Dh}),
obtaining
\begin{equation}
m \propto (b^*)^{-d} \frac{h(b^*)}{h_0} \left[
\frac{h(b^*)}{g(b^*)}\right]^{p_h} \sim (b^*)^{-d} \frac{1}{h_0} g(b^*)^{-p_h}
\sim (b^*)^{-d} \frac{1}{h_0} (\log b^*)^{ r p_h}\,,
\label{eq:deltaA}
\end{equation}
where $h(b^*)=1$
\begin{equation}
b^* \sim h_0^{-2/(d+2+2 c)} (\log h_0)^{-2 x /(d+2+2 c)}\,.
\end{equation}
By matching the l.h.s. and r.h.s. logarithms of equations (\ref{eq:mh}) and
(\ref{eq:deltaA}), respectively, we obtain the relation~(\ref{eq:deltaC}).

Remember that $2 c =-\eta$, $\nu=1/(\sigma+a)$ and $\hat{\nu}=-p \nu$.


\ukrainianpart

\title{Перегляд (логарифмічних) співвідношень скейлінгу з використанням ренормгрупи}
\author{Х.Х. Руіс-Лоренсо\refaddr{label1,label2,label3}}
\addresses{
  \addr{label1} Фізичний факультет, Університет Екстремадури, 06071 Бадахос, Іспанія
  \addr{label2} Інститут передових наукових обчислень (ICCAEx), Університет Екстремадури, 06071 м. Бадахос, Іспанія
  \addr{label3} Інститут біообчислень і фізики складних систем (BIFI), м. Сарагоса, Іспанія}

\makeukrtitle

\begin{abstract}
Ми явно обчислюємо критичні показники, пов'язані з логарифмічними поправками, виходячи з рівнянь ренормгрупи і середньопольової поведінки
для широкого класу моделей як при вищій критичній вимірності (для коротко- і далекосяжних $\phi^n$-теорій), так і нижче від неї.
Це дозволяє нам перевірити співвідношення скейлінгу, що пов'язують  критичні показники, аналізуючи комплексні сингулярності
(нулі Лі-Янга і Фішера) цих моделей.
Окрім того, ми запропонували явний метод для обчислення показника $\hat{\qq}$ [означеного як $\xi\sim L (\log L)^{\hat{\qq}}$]
і, накінець, ми отримали нове виведення закона скейлінгу, пов'язаного з цим показником.

\keywords ренормгрупа, скейлінг, логарифми, середнє поле
\end{abstract}

\end{document}